# Power Efficient Communication for Low Signal to Noise Ratio Optical Links

**RAVIKIRAN KAKARLA,**[1†] **MIKAEL MAZUR,**[2] **JOCHEN SCHRÖDER,**[1] **PETER A. ANDREKSON**[1*]

[1]*Department of Microtechnology and Nanoscience, Chalmers University of Technology, Gothenburg, 412-96, Sweden*
[2] *Nokia Bell labs, 600 Mountain Ave, New Providence, NJ 07974, USA.*
[†] *Currently at Optoelectronic research center, University of Southampton, UK*
*\*Corresponding author: peter.andrekson@chalmers.se*

**Abstract:** Receiver sensitivity is a particularly important metric in optical communication links operating at low signal to noise ratios (SNRs), for example in deep-space communication, since it directly limits the maximum achievable reach and data rate. Pulse position modulation (PPM) with direct detection photon counting detectors are the most power efficient solution known, however, the sensitivity gain comes at the expense of reduced spectral efficiency. We show that quadrature phase-shift keying (QPSK) modulation with a phase-sensitive ultralow noise pre-amplified coherent receiver outperforms other well-known power efficient multi-dimensional coherent modulation formats, while simultaneously having higher spectral efficiency. It also results in better sensitivity than PPM for orders up to 64 with ideal direct detection using photon counting receivers. This is because of the bit error rate characteristics favoring the QPSK format when forward error correction with a large overhead is considered.

## 1. Introduction

Optical receivers operating at low power and low signal-to-noise ratio (SNR) are highly desirable in applications such as deep-space laser communications, light detection and ranging (LIDAR), spectroscopy, quantum key distribution (QKD) and laser-interferometry-based gravitational wave observation [1]–[4]. The ability to detect optical signals with sufficient fidelity is fundamentally limited by the unavoidable vacuum noise. In addition, there exist sources of excess noise including background radiation, detectors with less than ideal quantum efficiency, and amplified spontaneous emission (ASE) if optically pre-amplified receivers are used.

In communications, in addition to the SNR and the detection technique, the choice of signal modulation format plays an important role in determining the link sensitivity. There is always a trade-off between spectral efficiency (SE, expressed as bits/(s.Hz)) and power efficiency (expressed e.g. as photons/bit, PPB) being dictated by the chosen modulation format. [5]. For example, the pulse-position modulation (PPM) format provides excellent receiver sensitivity, defined as the minimum average power needed to recover the information sent without error, often expressed as the PPB at the expense of low SE. This is not necessarily a concern if the target data rate is modest, but because of the limited bandwidth of the analog hardware (modulators, receivers, electronics), this will become a concern at sufficiently high bit rates.

In optical space communication, the data rate and communication reach of a link are limited by the available transmitter power, the receiver sensitivity and the channel loss as dictated by beam diffraction, optics losses, pointing losses etc. While the transmitted power and link loss are primarily limited by engineering capabilities, the receiver sensitivity is fundamentally limited by noise. In this paper we focus on these fundamental limitations with the aim to find the most suitable choice of modulation format and receiver implementation targeting the best possible sensitivity over a range of received powers.

It is well known that PPM along with photon counting detection provides the lowest possible sensitivity [5]. Drawbacks with this approach are the need to cool the receiver to cryogenic temperatures and the non-ideal detection quantum efficiency (degrading the sensitivity) especially for higher bandwidth detectors. Moreover as PPM spreads the energy over a broader bandwidth, it is spectrally inefficient and due to hardware limitations, in particular the receiver bandwidth, typical achievable data rates (i.e. without significant sensitivity degradation) are limited to well below 1 Gb/s in practice. An alternative approach is to use spectrally efficient modulation formats along with pre-amplified coherent receivers, as is widely used in optical fiber communication, supporting very high data rates. When used together with ultra-low-noise optical pre-amplifiers this approach enables extremely high sensitivities. Recently, we reported a record sensitivity of 1 photon per information bit (at BER < $10^{-6}$) using a phase-sensitive optical preamplifier (PSA) at 10.5 Gb/s [6], thus excelling in sensitivity and data rate simultaneously. This demonstration used 100% forward error correction (FEC) overhead redundancy for improved noise resilience as is commonly used in low SNR applications [7],[10].

We investigate here theoretically power-efficient optical modulation formats including several coherent formats with Gaussian noise statistics using erbium-doped fiber amplifier (EDFA) and PSA pre-amplification, as well M-PPM formats with direct-detection photon-counting with Poisson statistics. We find, both theoretically and experimentally, that at low SNRs, quadrature phase-shift keying (QPSK) is the most sensitive coherent format and it is not necessary to resort to advanced multi-dimensional multi-format modulation schemes, such as M-PPM+QPSK as adopted in some recent demonstrations [11][12]. Moreover, in comparison to PPM based photon-counting, PSA pre-amplified coherent detection of QPSK signals theoretically outperforms PPM up to extremely low received powers of around -85 dBm. PSA pre-amplification therefore has the potential to change the design trade-offs and significantly increase data rates and/or transmission reaches applicable to systems that operate at very low SNR.

## 2. Theoretical receiver sensitivities

PPM encodes information into the temporal position of an optical pulse relative to a symbol frame and is practically often used in combination with single-photon detectors. In contrast to coherent modulation formats which are widely applied in terrestrial optical fiber communication networks and are governed by Gaussian noise statistics, PPM with single-photon detection is governed by Poisson statistics. This can in the limit of low power and high PPM-order approach the ultimate sensitivity given by the so-called Gordon's capacity or Holevo limit [5]. In the following, we compare the capacity for coherent modulation and photon-counting PPM considering practical limitations on receiver bandwidths.

The capacity of M-PPM formats, governed by Poisson noise statistics, is given by [13] $C_{PPM} = B_{rec}(1-e^{-n_s})log_2(M)/M$ , where $n_s = \frac{P_{rec}}{h\nu B_{sym}} = \frac{MP_{rec}}{h\nu B_{rec}}$ is the average number of photons per $M$-PPM symbol, $P_{rec}$ is the average received power, $h$ is Planck's constant, and $\nu$ is the optical frequency of the wave. Here $B_{rec} = 1/\tau_s$ is the bandwidth of the receiver, with $\tau_s$ being the duration of the PPM time-slot which is $M$-times shorter than the symbol duration $t_{sym} = 1/B_{sym}$.

Coherent modulation formats are governed by Gaussian statistics and their capacity therefore derives from the well-known Shannon's capacity $C = B\log_2(1+SNR)$. Pre-amplifiers are employed partly to overcome the low quantum efficiency of the commercial coherent receivers, where the noise figure of the pre-amplifier plays a role similar to that of the quantum efficiency in a direct detection receiver [6]. For a pre-amplified receiver the signal-to-noise-ratio is given by $SNR = 2n_s / NF$ where $NF$ is the noise figure of the amplifier and $n_s = P_{rec} / h\nu B_{rec}$ is the average number of photons per symbol, with $B_{rec} = 1/t_{sym}$ being the bandwidth of the receiver. The capacity thus becomes $C_{pre-amp} = B_{rec} \log_2\left(1 + \frac{2n_s}{NF}\right)$. For an ideal EDFA $NF = 2$. For the PSA which transmits a signal and its phase-conjugate copy as an idler [6] $NF = 1/2$ for each wave. Hence the capacity of a PSA pre-amplified receiver is $C_{PSA} = B_{rec} \log_2(1 + 4n_s)$. Note that since the PSA requires the transmission of both signal and idler waves, if one were to consider the optical channel bandwidth, the loss of bandwidth due to idler needs to be accounted for, which would result in a smaller capacity $C_{PSA} = \frac{B_{optical}}{2} \log_2(1 + 4n_s)$ where $B_{optical}$ is the optical channel bandwidth. However, as low SNR links are generally limited by the receiver bandwidth and not the optical bandwidth and as the idler wave contains the same information as the signal and its noise is correlated to that of signal at the output of the PSA, it is not meaningful to also detect the idler. Therefore, the idler can be ignored in this analysis, and $B_{rec}$ remains the same for the EDFA and PSA cases. We note that the capacity and thus the sensitivity are in both cases limited by the available receiver bandwidth. A meaningful comparison should therefore take this bandwidth into consideration.

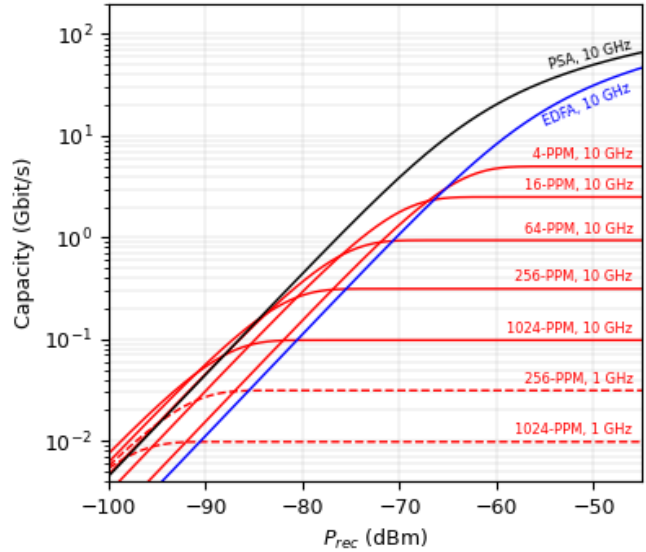

Fig. 1 Capacity vs. received power for coherent formats with pre-amplification governed by Gaussian noise statistics (PSA, EDFA) and PPM based on photon-counting detection governed by Poisson noise statistics. The theory assumes no implementation penalty and a receiver bandwidth of 10 GHz or 1 GHz (dashed lines), respectively.

Figure 1 shows a comparison of the capacity of coherent modulation formats with ideal EDFA and PSA pre-amplified receivers and for PPM with ideal photon-counting detection as a function of received average power. Here, we assume that there are no implementation penalties. We note that the PSA significantly outperforms the EDFA, especially at low power. In addition, the PSA dramatically changes the power where coherent formats outperform PPM. For the same 10 GHz receiver bandwidth, PSA amplified coherent modulation formats offer higher capacity than PPM formats up to 64-PPM. 256-PPM outperforms the PSA below a received power of -85 dBm. This relative comparison is, of course, valid for any chosen bandwidth. In comparison the EDFA only outperforms PPM at received powers above –65 dBm. Moreover, 10 GHz bandwidth is an optimistic estimate for the bandwidth of high quantum efficiency photon-counting detectors and current state-of-the-art devices at 1 GHz (dashed lines) have low quantum efficiencies [14]–[16] resulting in typical PPM implementation penalties of approximately 10 dB [9][17][18]. In contrast, we have shown that the implementation penalty of PSA pre-amplification with coherent modulation is about 3 dB [6].

While the discussion above highlights the advantages of PSA pre-amplified coherent detection for low SNR communications, it is, to date, unclear which is the best modulation format with coherent detection. As mentioned earlier, large FEC overheads are frequently used in communication links that operate at low SNR. It is thus convenient to introduce the information data rate R, and a raw bit rate, $R_t = R/k$ where k is the code rate (100% overhead corresponds to k = 0.5). In addition, we make the distinction of the photons per bit, with respect to raw transmitted (pre-FEC) bits ($PPB_t$) and actual (post-FEC) information bits (PPB), $PPB_t = PPB*k$. In recent experiments, SNRs corresponding to pre-FEC bit-error rate (BER) of 10-15% with 100 % overhead FEC have proven to be a suitable choice due to practical limitations from e.g. laser linewidths, digital signal processing considerations and FEC performance [19][20]. Modulation formats that are more power efficient than QPSK, such as polarization switched PS-QPSK [21], 3-PSK and hybrid formats that combine PPM and QPSK have shown better sensitivity than QPSK at BER = $10^{-3}$ as demonstrated in [22], [23]. On the other hand, we recently demonstrated a record 1 PPB 'black-box' sensitivity at 10 Gb/s using QPSK-only modulation with a PSA pre-amplified intradyne coherent receiver, with a NF =1.2 dB and a FEC with k=0.5

In the following, we will theoretically and experimentally compare the performance of the well-known power efficient modulation formats BPSK/QPSK, 3-PSK, M-PPM and M-PPM+QPSK [21], [23], [24] with PSA pre-amplified coherent detection and investigate the best possible format for a given received SNR.

In PPM, information is encoded in the position of a pulse among M slots, resulting in $\log_2 M$ bits/symbol. At the same time, the SE is reduced to at best $(\log_2 M)/M$ bits/s/Hz (or equivalently bits/symbol). The choice of M in PPM therefore corresponds to a SE versus sensitivity trade-off. In M-PPM+QPSK, the pulse in a M-PPM symbol is modulated with QPSK data, resulting in bits/symbol and a maximum SE of bits/s/Hz. The BER of M-PPM with QPSK is [22],

$$BER_{M-PPM+QPSK} = \frac{SER_{M-PPM}\left(1+\frac{M}{2(M-1)}\log_2 M\right)+(1-SER_{M-PPM})2BER_{QPSK}}{\log_2 M + 2} \quad (1)$$

where $SER_{M-PPM}$ is the symbol error rate of M-PPM and $BER_{QPSK}$ is the BER of QPSK. The expression for $SER_{M-PPM}$ was derived in [25]. The first term in the numerator corresponds to the bit errors caused by incorrectly detecting the PPM symbol and the second term is due to incorrectly detecting QPSK symbol when the PPM symbol is correctly detected.

For a pre-amplified coherent receiver, the conversion between SNR and PPB is given as follows. The SNR per symbol is $SNR_{sym} = 2n_s/NF$ where $n_s$ is the number of received photons per symbol. With N being the number of bits per symbol, $SNR_{bit} = SNR_{sym}/N$ and photons per bit, $PPB = n_s/N$, we obtain $PPB = NF(SNR_{bit})/2$.

Fig. 2b shows the theoretical BER (without FEC) versus $PPB_t$ (converted from SNRs) for M-PPM, M-PPM+QPSK, 3-PSK and QPSK with a PSA (NF = 0 dB) pre-amplified coherent receiver. We studied PPM with M=16, 64 and 128, corresponding to three distinct cases of each format. The two horizontal lines indicate BER =14%, the cut off pre-FEC BER for an FEC with about 100% OH, widely used in low SNR space communications [8], [26], and BER = $10^{-3}$), allowing low overhead (~7%) hard-decision FEC. At relatively high power, higher-order PPM+QPSK exhibits an advantage over both coherent PPM and QPSK. However, at low power (where BER > 8%) QPSK outperforms coherently detected M-PPM and M-PPM+QPSK. QPSK also outperforms 3-PSK over the considered range and 3-PSK only offers an advantage at higher SNRs. Similarly, as polarization-switched-QPSK performs the same as 2-PPM+QPSK it was not included in the graph.

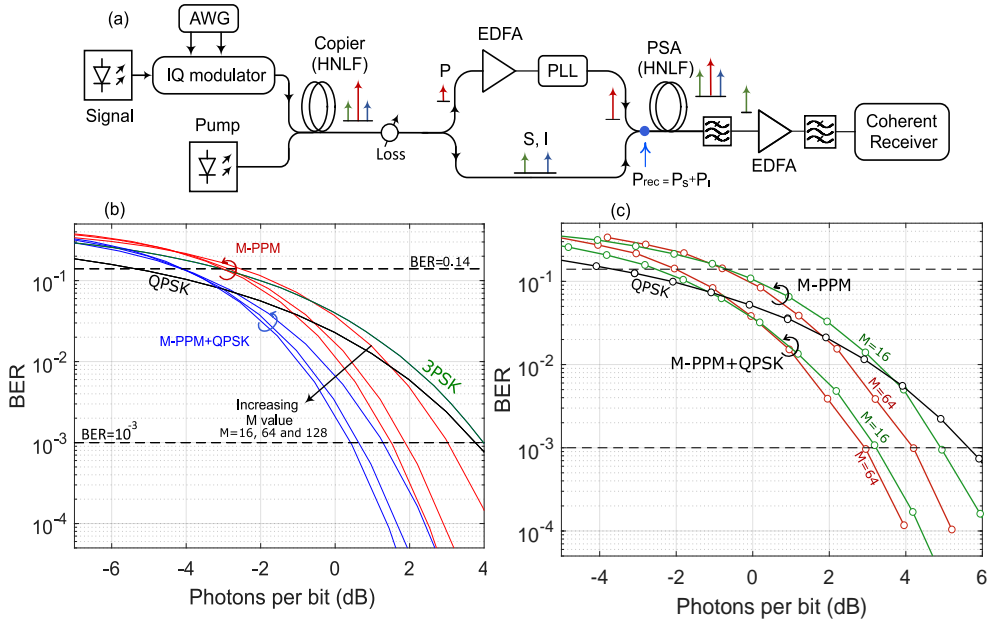

Fig. 2 (a) Experimental setup. (b) Theoretical plots of BER performance versus PPB of QPSK, 3-PSK, M-PPM, M-PPMs with QPSK, all using an ideal PSA pre-amplified coherent receiver. M=16, 64 and 128 (c) Experimental results using a 1.3 dB NF PSA pre-amplified coherent receiver with QPSK, M-PPM, M-PPMs with QPSK for M=16 and 64.

## 3. Experimental evaluation

To evaluate the theoretical results under realistic experimental conditions including implementation penalties we performed experiments with the PSA setup shown in Fig. 2a. A transmitter signal laser was modulated using 16-, 64- PPM and PPM+QPSK formats as well as the QPSK format, driven by an arbitrary waveform generator (AWG). The PPM slot rate and QPSK symbol rate were chosen to be the same at 10 GHz such that all signals had the same bandwidth. The modulated signal was then combined with a CW-pump in a copier stage where a conjugated signal (idler) was generated by four-wave mixing in a highly nonlinear fiber (HNLF). The waves were then sent to the PSA (consisting of cascaded HNLF spools) where the signal and idler were amplified with a gain of 21 dB and a NF =1.3 dB. The received power is sum of signal and idler powers measured at the point 'P$_{rec}$' indicated in Fig. 2 (a). The pump was used for both the copier and the PSA and its power was not taken into account for this comparison as it can be regenerated from ultra-low input powers using optical injection locking [27]. The amplified signal after the PSA was filtered, amplified further, and detected using a coherent receiver. A part of the signal laser was used as a local oscillator (LO) to perform self-homodyne detection and sampled using a real-time oscilloscope at 50GS/s. The data was processed using data-aided digital signal processing (DSP) for PPM and PPM+QPSK formats and with regular DSP for QPSK including blind equalization and phase estimation [28]. The experimental results are shown in Fig. 2c. At BER=$10^{-3}$, 64- performs better than 16- for both PPM and PPM-QPSK, and better than QPSK. At low received power, 16- performs better than 64- for both PPM and PPM-QPSK, in agreement with the theoretical plots, and hence it is better to operate at lower order PPMs with a better SE. We note that the SE, expressed in bits/(s·Hz) and accounting for a 100% OH, of the evaluated formats are: 1 (QPSK), 0.75 (3PSK), 0.125 (16-PPM), 0.047 (64-PPM), 0.19 (16-PPM+QPSK) and 0.062 (64-PPM+QPSK). The experimental results also confirm that QPSK outperforms all other coherently detected formats at low powers, *offering simultaneously the best sensitivity and highest SE*. Experimentally, we observe PPB = -3.7 dB (0.4 in linear units) at BER = 14%, which corresponds to 0.8 PPB post-FEC sensitivity, and which is 1.2 dB better than 16-PPM+QPSK. The sensitivities of all formats are summarized in Table 1. The implementation penalties (with the NF of the PSA contributing by 1.3 dB) range from 1.7 dB for QPSK (BER=14%) to 2.3 dB for 64-PPM+QPSK (BER = $10^{-3}$). We also note that, with the assumption that an FEC with 100% OH is capable of reducing a 14% BER to below $10^{-6}$ post-FEC, it results in an experimental sensitivity advantage over the case with pre-FEC BER =$10^{-3}$ in the range of 1.9 dB (64-PPM+QPSK) to 6.3 dB (QPSK). This clearly justifies the use of a large FEC overhead.

Table 1. Theoretical and experimental sensitivities in PPB (dB) with different modulation formats using PSA a pre-amplified coherent receiver at different BER. The best choice at each pre-FEC BER is indicated in bold.

|  | BER=$10^{-3}$ | | | BER= 0.14 | | |
|---|---|---|---|---|---|---|
| PPM order | 16 | 64 | QPSK | 16 | 64 | QPSK |
| PPM experiment | 4.9 | 4.2 |  | -0.6 | -0.6 |  |
| (theory) | (3.1) | (1.9) | 5.6 | (-2.4) | (-2.8) | **-3.7** |
| PPM+QPSK experiment | 3.2 | **2.9** | (3.8) | -2.5 | -2 | **(-5.4)** |
| (theory) | (1.3) | **(0.7)** |  | (-4.1) | (-4.1) |  |

## 4. Conclusions

In conclusion, we theoretically and experimentally evaluated various coherent modulation formats with a PSA pre-amplified coherent receiver, and among the evaluated formats, the best sensitivity at a pre-FEC BER of 14% is experimentally achieved with QPSK (0.4 PPB corresponding to an ideal post 100%-OH-FEC sensitivity of 0.8 PPB), while the best sensitivity at a pre-FEC BER = $10^{-3}$, was obtained with 64-PPM-QPSK (1.9 PPB assuming a 7% FEC overhead in this case).

We note that our investigation only included the limitations of the receiver sensitivity for the optical field that reaches the receiver. A similarly important practical aspect in free-space and deep-space communications is the amount of light that can be coupled to the receiver, which is determined by the effective aperture of the capturing optics and the size of the receiver. Techniques for maximizing the amount of captured light, include for example multi-mode receivers [30], large capturing lenses or arrangements of multiple small apertures together with analog and digital coherent combining [29][10] and adaptive optics techniques. Different approaches may pose significant challenges to the different modulation and amplification schemes and is the topic of ongoing research, but beyond the scope of this article.

**Funding**

Funding was provided by the Swedish Research Council (VR) under project grant VR-2015-00535 and by the KA Wallenberg Foundation.

**Acknowledgments**

The authors would like to thank Zonglong He, Kovendhan Vijayan and Rasmus Larsson for assistance in the laboratory, Roland Ryf with assistance for the DSP development for signal recovery, and Ali Mirani and Magnus Karlsson for valuable discussions.

**Disclosures**

The authors declare no conflicts of interest.